\documentclass[useAMS,usenatbib]{mn2e}

\usepackage{epsfig}
\usepackage{amsmath}
\usepackage{amssymb}
\usepackage{natbib}
\usepackage{times}
\usepackage{threeparttable} 
\usepackage{booktabs}

\usepackage{epstopdf}

\usepackage{multicol}

\newcommand{\swift}{\textit{Swift}}

\newcommand{\xmm}{\textit{XMM-Newton}}

\newcommand{\nh}{\mathrm{cm}^{-2}}

\newcommand{\vsource}{V0332+53}
\newcommand{\usource }{4U 0115+63}

\usepackage[dvips]{color}

\def \mnras {MNRAS}
\def \apj {ApJ}
\def \apjs {ApJS}
\def \apss {Ap\&SS}
\def \apjl {ApJL}
\def \aap {A\&A}

\def \pasj {PASJ}

\title[Sub-luminous X-ray states in V0332+53 and 4U 0115+63]{Meta-stable low-level accretion rate states or neutron star crust cooling in the Be/X-ray transients V0332+53 and 4U 0115+63}

\author[Wijnands et al.]{
\parbox[t]{\textwidth}{
\raggedright
R. Wijnands$^{1}$\thanks{r.a.d.wijnands@uva.nl} \&
N. Degenaar$^{1,2}$
}
\vspace{6pt}\\
$^{1}$Anton Pannekoek Institute for Astronomy, 
University of Amsterdam,
Postbus 94249, 1090 GE Amsterdam, The Netherlands\\
$^{2}$Institute of Astronomy, University of Cambridge, Madingley Road, Cambridge, CB3 OHA, UK
}

\begin{document}


\pagerange{\pageref{firstpage}--\pageref{lastpage}} \pubyear{0000}

\maketitle

\label{firstpage}

\begin{abstract}The Be/X-ray transients \vsource\ and \usource\ exhibited giant, type-II outbursts in 2015. Here we present {\it Swift}/XRT follow-up observations at the end of those outbursts. Surprisingly, the sources did not decay back to their known quiescent levels but stalled at a (slowly decaying) meta-stable state with luminosities a factor $\sim$10 above that observed in quiescence. The spectra in these states are considerably softer than the outburst spectra and appear to soften in time when the luminosity decreases. The physical mechanism behind these meta-stable states is unclear and they could be due to low-level accretion (either directly onto the neutron stars or onto their magnetospheres) or due to cooling of the accretion-heated neutron star crusts. Based on the spectra, the slowly decreasing luminosities, and the spectral softening, we favour the crust cooling hypothesis but we cannot exclude the accretion scenarios. On top of this meta-stable state, weak accretion events were observed that occurred at periastron passage and may thus be related to regular type-I outbursts.
\end{abstract}

\begin{keywords}
accretion, accretion discs -- binaries: close - pulsars: individual: V0332+53, 4U 0115+63 - stars: neutron - X-rays: binaries  
\end{keywords}

\section{Introduction}

In a Be/X-ray transient, a strong magnetic-field ($B$$\sim$$10^{12-14}$~G) neutron star is accreting matter from a Be star \citep[see][for a review]{2011Ap&SS.332....1R}. Typically the
orbit is very eccentric and
matter is accreted only during outburst episodes. Two types of outbursts have been identified in those
systems \citep[e.g.,][]{1986ApJ...308..669S}. The
so-called normal, type-I outbursts occur because the neutron star passes through
periastron and transits through the decretion disk of the Be
star, allowing it to accrete some of this matter \citep[see, e.g., the model of][]{2001A&A...377..161O}. Type-I outbursts typically peak at an X-ray luminosity of $L_{\mathrm{X}}$$\sim$$10^{36-37}$ erg s$^{-1}$. In addition, some transients also exhibit very
bright, giant type-II outbursts with typical luminosities that are an order of magnitude higher than during the
type-I outbursts, reaching the Eddington limit for a neutron star ($L_{\mathrm{Edd}}$$\sim$$2\times10^{38}$ erg s$^{-1}$). Those outbursts can also last much longer
than an orbital period. What causes those type-II
outbursts is still not understood but one possibility is that a misaligned (with the orbital plane) decretion disk exist around the Be star that becomes sufficiently warped and eccentric in time for the neutron star to capture a large amount of matter near periastron passage \citep[e.g.,][]{2013PASJ...65...41O,2014ApJ...790L..34M}.

Be/X-ray transients are intensively studied during their outbursts, but little is known
about their behaviour when $L_{\mathrm{X}}$$<$$10^{35-36}$ erg
s$^{-1}$. Some systems have been detected at levels of $L_{\mathrm{X}}$$\sim$$10^{34-35}$ erg s$^{-1}$ \citep[e.g.,][]{1991ApJ...369..490M,2002ApJ...580..389C,2007ApJ...658..514R}. This luminosity is likely caused by low-level
accretion onto the neutron star or onto its magnetosphere. Exactly how this occurs is
not well understood and it could strongly depend on the rotation period of the neutron star (and to a lesser degree on its magnetic
field strength). In systems with a slow spin period (hundreds of seconds), direct accretion onto the neutron star may still occur but faster spinning neutron stars (spin periods
of seconds) should instead be
in the so-called propeller regime. Matter can then no longer 
accrete onto the surface, but might instead be propelled away
by the rotating neutron star magnetic field \citep[e.g.,][]{1975A&A....39..185I,1986ApJ...308..669S}. Some accretion onto the surface might still be possible if matter ``leaks" through the magnetosphere due to some instability \citep[e.g.,][]{1977ApJ...215..897E,2001A&A...375..944I,2004ApJ...616L.151R,2014MNRAS.441...86L}.

Several Be/X-ray transients
have been detected at very low (quiescent) X-ray luminosities of $L_{\mathrm{X}}$$\sim$$10^{32-34}$ erg s$^{-1}$ \citep[e.g.,][]{1987ApJ...312..755M,2001ApJ...555..967R,2002ApJ...580..389C,2014MNRAS.445.1314R}. The origin of this faint emission is unclear. Proposed
mechanisms include very low-level accretion onto the neutron star magnetic poles \citep[strong evidence exist for this scenario for A0535+26; e.g.][]{2013ApJ...770...19R,2014A&A...561A..96D}, although here it is also unclear how matter can leak through the magnetosphere at such low rates.  It could also be due to emission from accretion onto the magnetosphere. Alternatively, the X-rays could be due to emission from the
neutron star surface that is cooling after being heated during the
preceding outburst \citep[e.g.,][]{brown1998,2002ApJ...580..389C,wijnands2012}.

To investigate the low luminosity behaviour of Be/X-ray transients, we have been following some of these systems with
{\it Swift} after the end of their outbursts to study the transition into the propeller regime \citep[][PI: Tsygankov]{sergeisasha2}, and to determine if the neutron star crusts might be heated due to the accretion of matter and can be observed to cool, as is the case for low-magnetic field neutron stars (PI: Wijnands). Here we report on the first results of the latter study using our {\it Swift} observations of \vsource\, and
\usource. Both sources have similar spin periods (4.4 and 3.6 s), orbital periods (33.8 and 24.3 days) and surface magnetic field strengths ($3\times10^{12}$ and $1.3\times10^{12}$ Gauss) \citep[see][for additional background information on these
two sources]{2005A&AT...24..151R,2015arXiv150900230C}. Both sources exhibited a giant, type-II outburst in 2015
\citep[][]{2015ATel.7685....1N,2015ATel.8179....1N,2015ATel.7822....1D,2015arXiv150904490D} and they are excellent candidates for this study, because they are very faint in their
quiescent state \citep[{$\sim3\times 10^{32}$ and $\sim6\times 10^{32}$ erg s$^{-1}$;}][]{2001ApJ...561..924C,sergeisasha}.

\section{Observations, analysis and results \label{selectionresults}}

\begin{figure}
 \begin{center}
\includegraphics[width=1.01\columnwidth]{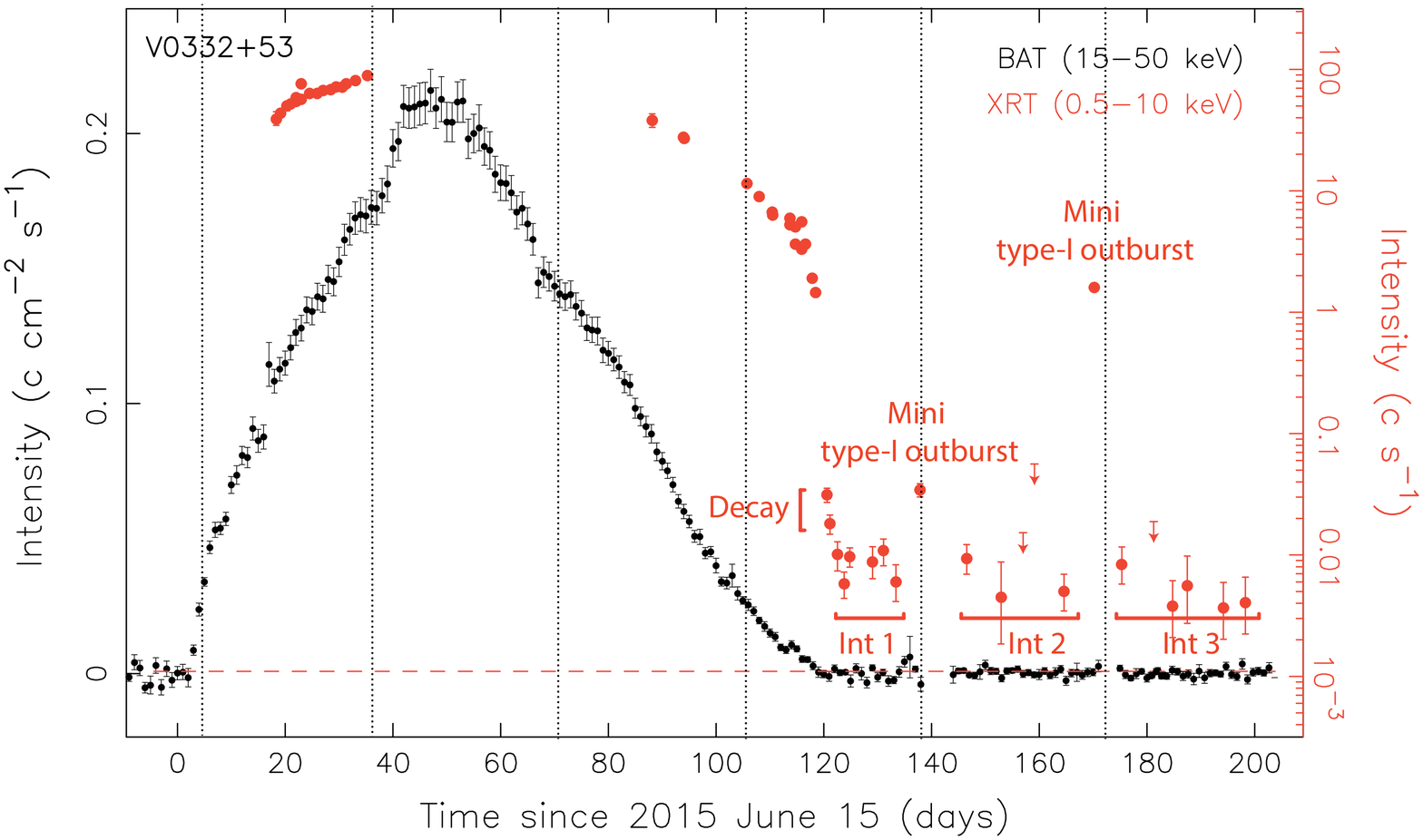}\vspace{+0.30cm}
\includegraphics[width=1.01\columnwidth]{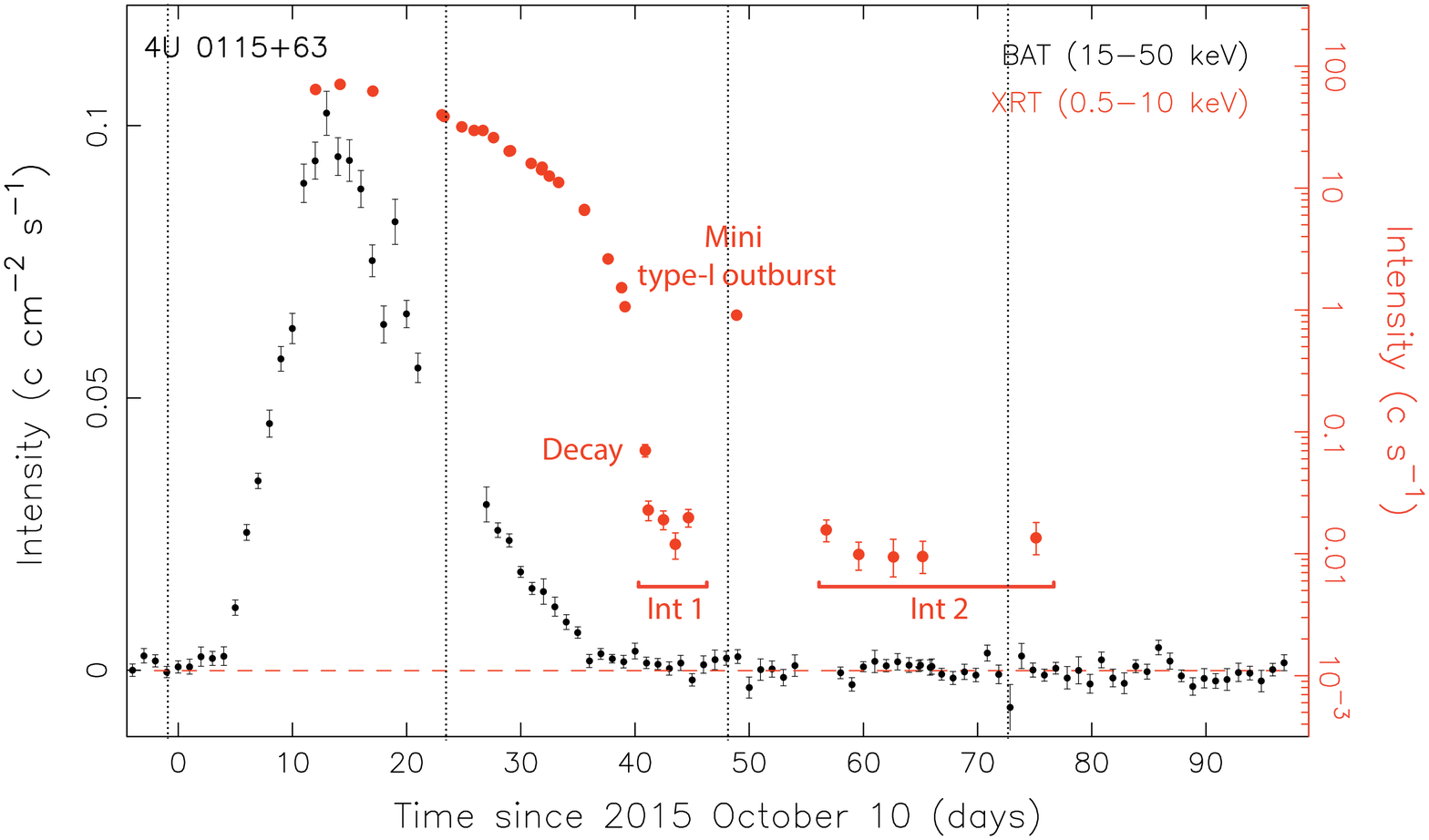}\vspace{-0.15cm}
    \end{center}
 \caption[]{{\it Swift} light curves of \vsource\ (top) and \usource\ (bottom) of the giant 2015 outbursts 
(BAT: black, 1-day bins; XRT: red, binned per observation). Red dashed horizontal lines indicate quiescent detections with \xmm\ reported by \citet{sergeisasha}, which we converted to XRT count rate equivalents using \textsc{webpimms} using the spectral parameters given in that work. Vertical dotted lines indicate periastron passages.
 }
 \label{lcsXRT}
\end{figure}

The 2015 type-II outbursts of both sources were covered by the {\it Swift}/BAT transient monitoring program \citep[][]{2013ApJS..209...14K}\footnote{http://swift.gsfc.nasa.gov/results/transients/}, and densely monitored with {\it Swift}/XRT (see Figure~\ref{lcsXRT}). We extracted the
XRT light curves using the online XRT data products tool \citep[][]{evans2007,evans2009},\footnote{Available at http://www.swift.ac.uk/user$\_$objects/} which is shown in
Figure~\ref{lcsXRT} (red). Detailed analysis of the decay phase is presented by \citet{sergeisasha2}, who focusses on the transition from
the direct accretion regime into the propeller state. This
transition likely occurred when the decay rate
accelerated, causing the sources to decrease rapidly in X-ray
luminosity (Figure~\ref{lcsXRT}).
However, the sources did not decay all the way into quiescence but
suddenly stalled at a luminosity level a factor of $\sim$10 brighter than
what has been observed previously in quiescence \citep[][Figure~\ref{lcsXRT}]{2002ApJ...580..389C,sergeisasha}. As can be seen from
Figure~\ref{lcsXRT}, it is clear that both sources stayed in
this state for at least 1--2 months. 
Nevertheless, for both sources
the count rate is slowly decreasing in this time interval.

From Figure~\ref{lcsXRT}, it can also been seen that on top of this
low luminosity state, the sources exhibited rebrightening episodes
during which the count rate increased by a factor of a few (for
\vsource\ around day 140) or even by approximately 2 orders of
magnitude (around day 170 for \vsource\ and around day 50 for \usource). All flares must have lasted $\lesssim$2 weeks. None of the flares were
seen in the BAT light curves. The 1-day sensitivity (for 15-50 keV) of the BAT is 5.3 mCrab \citep[$\sim$$7\times10^{-11}$ erg cm$^{-2}$ s$^{-1}$;][]{2013ApJS..209...14K}, limiting the peak brightness of these episodes to $L_{\mathrm{X}}$$\lesssim$$4\times10^{35}$ erg s$^{-1}$ (assuming a distance of 7 kpc for both sources). Those rebrightening events occurred at times of periastron passage of the neutron star (based on the ephemerides of \citet{2015arXiv150904490D} and \cite{2010MNRAS.406.2663R}; dotted lines in Figure~\ref{lcsXRT}). Therefore, these are likely related to type-I outbursts, albeit much fainter. Hence we call these rebrightening events ``mini type-I outbursts''.  Figure~\ref{lcsXRT} shows that the
rebrightening episodes did not significantly affect the
underlying low luminosity state (hereafter ``the meta-stable state''): both sources were observable at very
similar count rates before and after these events.

The {\it
Swift}/XRT was used in Photon Counting (PC) mode, which provides a time resolution of
2.5~s. This is insufficient to search for pulsations, since the pulse period of \vsource\, is 4.4~s and 3.6~s for
\usource. Furthermore, the XRT count rates of both sources are very low, inhibiting any sensitive search for pulsations even when the Window Timing (WT) mode would be used.

\begin{figure}
 \begin{center}
\includegraphics[width=\columnwidth]{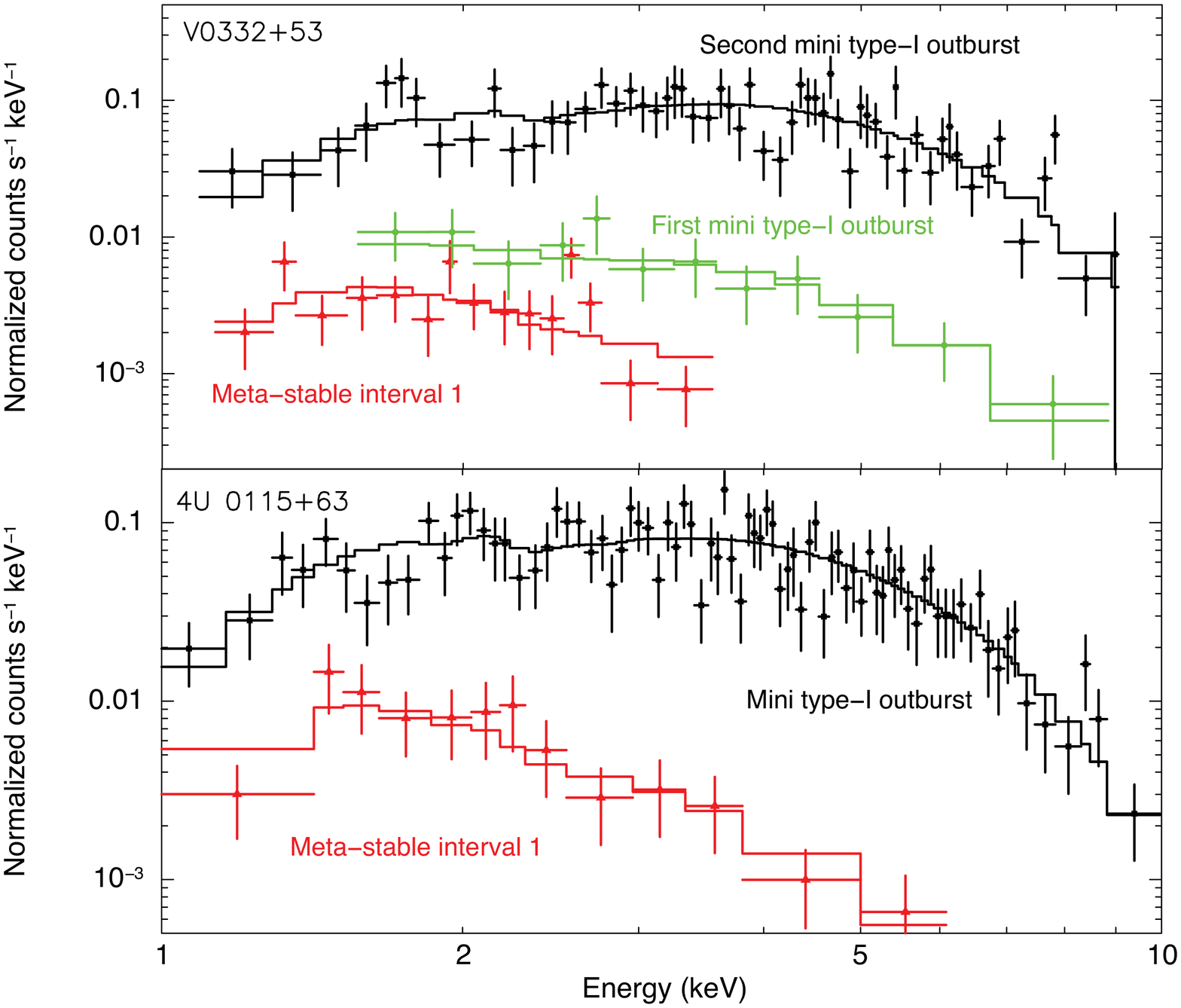}\vspace{-0.15cm}
    \end{center}
 \caption[]{Representative {\it Swift}/XRT spectra of \vsource\ (top) and \usource\ (bottom) fitted to a power-law model (rebinned for visual clarity). }
 \label{spectra}
\end{figure}

We investigated the spectral properties of the outburst decay, meta-stable state, and
flare episodes of both sources. To this end we obtained {\it Swift}/XRT data from the
HEASARC archive up to 2015 December (obsID 312930[31--52] for \vsource\, and 311720[31--41] for \usource). Typical exposure times were 1--4~ks per observation. 
The data were analysed with tools
incorporated within \textsc{heasoft} v. 6.16. After reprocessing
the raw data with the \textsc{xrtpipeline}, we extracted source and
background spectra using \textsc{XSelect}. Source counts were obtained
from a circular region with a radius of 15 pixels, and background
counts from a surrounding annulus with an inner-outer radius of
60--110 pixels. Exposuremaps were used to create ancillary response files using
\textsc{xrtmkarf} and the latest response matrix files (v. 15)
were used. 

Due to the very low count rates detected during the meta-stable state, we
combined several subsequent observations to get good enough
statistics to fit the spectra (Figure~\ref{lcsXRT} and Table~\ref{spectralresults}). Separate spectra could be extracted for the rebrigthening episodes. The spectral data were fitted in the 0.5--10 keV energy range using \textsc{XSpec} v. 12.8 \citep[][]{xspec}.
{Given the low number of counts per spectrum ($\lesssim$150), we binned the spectra to a minimum of 1 count per bin (with \textsc{grppha}) and used C-statistics. For each fit we performed Monte Carlo simulations with the \textsc{goodness} command in \textsc{XSpec} (using $10^4$ realizations). The percentage of those simulations having a larger C-statistic than the data is reported as the ``goodness'' in Table~\ref{spectralresults}.}

We fitted the spectral data with two single component models: a
power-law model (\textsc{pegpwrlw}) and a black-body model
(\textsc{bbodyrad}). In all fitting we included absorption by the
interstellar medium (\textsc{tbabs}) with abundances set to
\textsc{wilm} and cross-sections to \textsc{vern} \citep[][]{verner1996,wilms2000}. For the column
density we used the Galactic values: $N_H$$=$$ 7\times10^{21}~\nh$ for \vsource, and $N_H$$= $$9\times10^{21}~\nh$ for
\usource\ \citep[][]{2005A&A...440..775K}. We adopted a distance of $D$$=$7~kpc for both sources \citep[see][]{1999MNRAS.307..695N,2001A&A...369..108N}.
For the
\textsc{bbodyrad} fits, we left the emitting radius as a free
parameter and determined the unabsorbed 0.5--10 keV flux
by using the \textsc{cflux} convolution model. 
In the \textsc{pegpwrlw} model we set the energy boundaries to 0.5 and 10
keV, so that the model normalization gives the unabsorbed flux in that band. The results of our spectral analysis are given in
Table~\ref{spectralresults}.  Representative spectra are shown in
Figure~\ref{spectra}.

Both sources are detected at $L_{\mathrm{X}}$$\sim$$10^{33-34}$ erg s$^{-1}$ during their meta-stable states, whereas during the mini type-I outbursts the luminosity is higher ($L_{\mathrm{X}}$$\sim$$10^{34-36}$ erg s$^{-1}$). 
The \textsc{pegpwrlw} fits suggest that the spectra of the meta-stable state are softer than that of the brightest mini type-I outbursts (Table~\ref{spectralresults}). 
The spectra of the meta-stable state can also be adequately described by a \textsc{bbodyrad} model with a temperature of $kT_{\mathrm{bb}}$$\sim$0.5--0.7~keV. 
These fits suggest, however, an emission radius that is smaller than the expected radius of a neutron star ($R_{\mathrm{bb}}$$\sim$0.3--0.6~km; Table~\ref{spectralresults}). This may indicate the presence of hot spots that could correspond to the magnetic poles of the neutron star.  Due to the limited data quality, we cannot statistically prefer one of the models over the other.

\begin{figure}
\begin{center}
\includegraphics[width=\columnwidth]{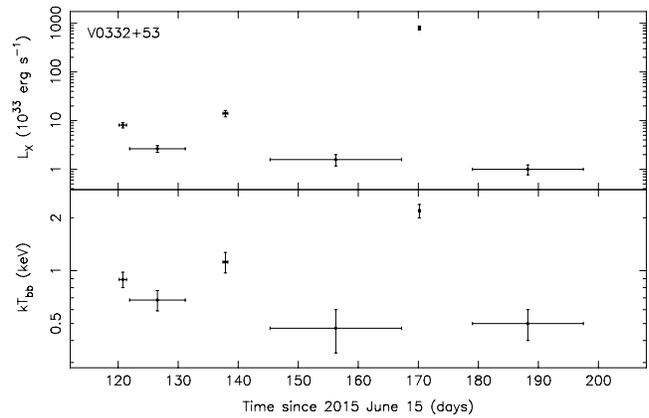}\hspace{0.5cm}
  \end{center}
\caption[]{Evolution of the black-body luminosity (top; 0.5-10 keV) and black-body temperature (bottom) for \vsource\ during its meta-stable state. }
\label{v0332BB}
\end{figure}

In the meta-stable state (ignoring the rebrightening events) the spectra seem to soften in time slightly for both sources, e.g., the photon indices increase from $\sim$1.2 to $\sim$2.8 for \vsource\ and $\sim$2.3 to $\sim$2.7 for \usource. Similarly, the temperatures in the black-body models decrease from $\sim$0.68 keV to $\sim$0.50 keV for \vsource\ (see Figure~\ref{v0332BB}) and from $\sim$0.66 keV to $\sim$0.54 keV in \usource.  
We note that in the \textsc{bbodyrad} model the emitting radius slightly varies and possible temperature changes may thus be entangled with possible changes in the hotspot size.

\begin{table*}
\caption{Results from our spectral analysis}
\begin{threeparttable}
\begin{tabular*}{1.04\textwidth}{@{\extracolsep{\fill}} rlccccccccc}
\hline
ObsID         &  Interval        &  $\Gamma$      & $F_{\mathrm{X}}$                &  $L_{\mathrm{X}}$  &  goodness &   $kT_{\mathrm{bb}}$ &   $R_{\mathrm{bb}}$     & $F_{\mathrm{X}}$            &  $L_{\mathrm{X}}$  &  goodness \\
              &                  &                 & (erg cm$^{-2}$ s$^{-1}$) & (erg s$^{-1})$ &  \% &  (keV) & (km) & (erg cm$^{-2}$ s$^{-1}$) & (erg s$^{-1})$ & \% \\
\hline 
312930+ & \multicolumn{10}{c}{V0332+53}\\  
     31--32 &  Decay           &  1.41$\pm$0.23  &   2.10$\pm$0.27           & 12.3$\pm$1.6     & 21.2 & 0.89$\pm$0.09     &  0.32$\pm$0.05             &   1.38$\pm$0.17           & 8.09$\pm$1.00          & 12.87 \\
     33--38  &  Interval I      &  1.17$\pm$0.33  &   1.06$\pm$0.26            &  6.21$\pm$1.53     & 38.47 & 0.68$\pm$0.09     &  0.31$\pm$0.06          &   0.45$\pm$0.07            & 2.64$\pm$0.41          & 29.70 \\
       40   &    Mini type-I         &  1.32$\pm$0.30  &   3.42$\pm$0..46           & 20.5$\pm$2.2    & 0.15 & 1.12$\pm$0.15    &  0.27$\pm$0.06  &  2.40$\pm$0.37    & 14.1$\pm$2.1 & 0.60 \\
    41--45   &  Interval II     &  2.60$\pm$0.94    &   0.57$^{+0.38}_{-0.14}$             &  3.34$^{+2.23}_{-0.82}$      & 60.46  & 0.47$\pm0.13$       &  0.51$^{+0.26}_{-0.09}$            &   0.27$\pm$0.07 &  1.58$\pm$0.41          & 62.11 \\
        46  &  Mini type-I     &  0.08$\pm$0.10  &  170$\pm$12.0             & 997$\pm$40       & 12.31 & 2.19$\pm$0.19     &  0.63$\pm$0.03           &  136.5$\pm$11.1     & 800$\pm$65    & 8.16 \\
   47--52    & Interval III     &   2.80$\pm$0.65           &     0.33$\pm$0.14                   &  1.93$\pm$0.83             & 2.57  &   0.50$\pm$0.10            &  0.35$^{+0.19}_{-0.08}$                     &     0.17$\pm$0.04                   &  1.00$\pm$ 0.23                 & 1.18 \\
\hline
311720+ & \multicolumn{10}{c}{4U 0115+63}\\ 
31  & Decay            &     1.66$\pm$0.25    &  4.32$\pm$0.55            & 25.3$\pm$3.3     & 39.22  &        0.90$\pm$0.10           & 0.46$\pm$0.05   &  3.00$\pm$0.43           & 17.6$\pm$2.5     & 39.13 \\
     32--35  & Interval I       &     2.29$\pm$0.33    &  2.61$\pm$0.41            & 15.3$\pm$2.4     & 0.02 &         0.66$\pm$0.07          &  0.60$\pm$0.06    &  1.45$\pm$0.19           & 8.50$\pm$1.12     & 0.19  \\
        36  & Mini type-I      &     0.32$\pm$0.08  &  85.9$\pm$4.6             & 504$\pm$27       & 0.87  &       1.84$\pm$0.10            &  0.58$\pm$0.02  &  67.3$\pm$4.0              & 395$\pm$23       & 0.58 \\
     37--41  & Interval II      &  2.65$\pm$0.30    &   1.11$\pm$0.19            & 6.51$\pm$1.11      & 30.72  &         0.54$\pm$0.05          &  0.57$\pm$0.05    &   0.58$\pm$0.07            & 3.40$\pm$0.41      & 12.90  \\     
\hline
\end{tabular*}
\label{spectralresults}
\begin{tablenotes}
\item[]Notes. -- $N_{\mathrm{H}}$ was fixed to $7\times10^{21}$ cm$^{-2}$ for \vsource\, and  $9\times10^{21}$ cm$^{-2}$ for \usource. $F_{\mathrm{X}}$ is the unabsorbed 0.5--10 keV flux in units of $10^{-12}$ erg s$^{-1}$ cm$^{-2}$. $L_{\mathrm{X}}$ is the corresponding luminosity in units of $10^{33}$ erg s$^{-1}$ assuming a distance of 7 kpc for both sources. {Errors represent 1$\sigma$ confidence levels. }
\end{tablenotes}
\end{threeparttable}
\end{table*}

\vspace{-0.2cm}
\section{Discussion}\label{sec:discuss}

We have presented the results of {\it Swift}/XRT monitoring observations of the Be/X-ray transients \vsource\ and \usource, obtained after the decay of their giant outbursts in 2015. Remarkably, both sources did not directly decay into quiescence, but stalled at a (slowly decaying) meta-stable plateau for $>$1--2 months that is $\sim$1 order of magnitude brighter than their known quiescent levels \citep[see Figure~\ref{lcsXRT};][]{2001ApJ...561..924C,sergeisasha}. 

Both sources exhibited brief ($<$2~weeks) rebrightening episodes during this meta-stable state coinciding with the neutron star being closest to the Be star (periastron passages). These were thus accretion events that are likely related to type-I outbursts albeit with a lower peak luminosity ($L_{\mathrm{X}}$$\sim$$5\times10^{35} - 10^{36}$~erg~s$^{-1}$). An event with very similar properties (i.e., outburst duration/fastness, peak luminosity, spectral shape) was observed for \usource\ by \citet{2001ApJ...561..924C}.  They proposed that such events are due to accretion in the transition regime between direct neutron star accretion in the bright outbursts (which have $L_{\mathrm{X}}$$>$$10^{36}$~erg~s$^{-1}$) and the magnetospheric accretion/propeller regime ($L_{\mathrm{X}}$$<$$10^{34}$~erg~s$^{-1}$). 

\citet{2001ApJ...561..924C} also found that in this transition regime, only a small variation in the accretion rate could produce very large luminosity fluctuations ($\sim$2 versus $>$250, respectively). Therefore, the mini type-I outbursts could only signify a small, temporarily  increase in the accretion rate down to the magnetosphere, just enough to penetrate it and produce a short-lived, weak outburst. It is unclear how and where such a density enhancement may be created, but the fact that they occur at periastron passage suggest that when the neutron star passes close to the Be star, the matter density near the neutron star passes the threshold for the system to enter this transitional regime. Detailed modelling of this regime is necessary to understand the physics behind these mini type-I outbursts.

Those small accretion events appeared not to alter the underlying meta-stable states. Fitting the X-ray spectra to a simple power-law or black-body model suggests that the final decay phases of the giant outbursts and the mini type-I outbursts are spectrally harder than the meta-stable state (Table~\ref{spectralresults}). It is clear that those events are due to accretion
onto the neutron star surface or its magnetosphere, but the origin of the X-ray emission in the meta-stable state is more uncertain. Here we consider three
models explaining this state: direct low level
accretion onto the neutron star magnetic poles, accretion down the magnetosphere, and cooling of an accretion heated neutron star.

\vspace{-0.3cm}
\subsection{Accretion down to the magnetospheric boundary \label{propeller}}

In this scenario, the matter in the accretion flow is halted at the magnetospheric boundary \cite[sometimes called magnetospheric accretion; e.g.,][]{1996ApJ...457L..31C,2001ApJ...561..924C} and no accretion onto the neutron star takes place (i.e., the propeller regime). 
\citet{2002ApJ...580..389C} calculated that the maximum bolometric luminosity emitted by this process should be $0.3 - 3\times 10^{34}$ erg s$^{-1}$ for \vsource\ and $0.6 - 6 \times 10^{33}$ for \usource\ \citep[scaled to 7 kpc; see also][for \usource]{2001ApJ...561..924C}. Those luminosities are remarkably close to what we observe in the meta-stable state, possibly indicating that we could indeed be observing accretion down to the magnetospheric boundary. However, \citet[][]{2001ApJ...561..924C} calculated bolometric luminosities and it is unclear how much of the radiation will be emitted in the X-ray band we used here.

Although \cite{2001ApJ...561..924C} did not elaborate on what kind of spectra one would expect for magnetospheric accretion, they argued that spectral differences between magnetospheric accretion and direct accretion onto the neutron star are to be expected. This conceivably could explain the difference between the spectra we observe during the meta-stable state and the other states. However, if indeed the meta-stable X-ray spectra are black-body like with a small emitting radius, this would strongly suggest it comes from the neutron star magnetic poles and therefore discounting magnetospheric accretion. Detecting relatively strong pulsations would also argue against magnetospheric accretion as it is expected to produce un-pulsed X-ray emission \citep{2001ApJ...561..924C}.

\vspace{-0.3cm}
\subsection{Direct accretion onto the neutron star magnetic poles}
In the direct accretion scenario, it is expected that
when matter falls onto the neutron star it is channeled by the magnetic
field to its magnetic poles, producing hot
spots. This could conceivably
produce spectra that are softer than
those observed when the sources are accreting at higher rates. However, it remains to be established whether the difference in spectra between
the meta-stable state and the very faint mini type-I outburst in \vsource\, can be
explained with only a factor of a few change in accretion rate; 
There are no calculations of the X-ray spectra expected from direct accretion of matter onto the neutron star magnetic poles at very low levels. 

Moreover, an important unresolved problem in the direct accretion scenario is how the matter eventually reaches the neutron star surface since the magnetic field would inhibit direct accretion when the accretion rate is very low \citep[i.e., the propeller regime; see e.g.,][]{1975A&A....39..185I,1986ApJ...308..669S}. However, the detection of pulsations {(at $\sim$103 s)} in A0535+26 at very low luminosities ($L_{\mathrm{X}}$$\sim$$10^{33}-10^{34}$~erg~s$^{-1}$; {indicating that the source was in the propeller regime despite its low spin period}) in combination with clear accretion signatures \citep[e.g., strong aperiodic variability in the light curve, detection of the source up to 100 keV;][]{2004NuPhS.132..476O,2005A&A...431..667M,2013ApJ...770...19R,2014A&A...561A..96D} demonstrated that direct accretion is possible in the propeller regime. Pulsations at similarly low luminosities have been seen in other systems too \citep[see Table 2 of][]{2014MNRAS.445.1314R}, but we argue in Section~\ref{seccool} that the presence of pulsations in itself does not exclude thermal emission from the neutron star surface due to cooling emission (i.e., not accretion) as the origin of the X-ray emission.

\vspace{-0.2cm}
\subsection{Cooling of an accretion-heated neutron star \label{seccool}}

The last scenario we consider is the one in which the neutron star crust is significantly heated during the bright type-II outbursts and cools once accretion has halted, until it re-establishes thermal equilibrium with the core. The same scenario has been proposed to explain the cooling curves observed in X-ray binaries containing low magnetic field ($B$$\lesssim$$10^{8}$~G) neutron stars. For such systems, the neutron star crust is
expected to be heated during accretion outbursts because it is
compressed by the matter accreting on its surface, causing a variety of nuclear
reactions (mainly pycnonuclear fusion) that occur deep in the
crust \citep[e.g.,][]{haensel1990a,haensel2008}: the ``deep crustal heating model'' \citep[][]{brown1998}. 

The magnetic field in these accreting neutron stars is expected to play no role and the whole crust is expected to be heated uniformly. This may not be true for the high magnetic
field neutron stars discussed here. During the bright
outbursts, the matter is preferable accreted at the magnetic poles. In first principles the matter is expected
to spread out over the rest of the surface very quickly, causing an
uniform compression of the whole crust and thus that a uniform heating from pycnonuclear
reactions. However, if it is possible to confine the
accreted matter to the poles without it spreading over the surface,
then only the crust at the poles is compressed. Eventually, the
matter has to spread out uniformly, but it is unclear how this would happen and if it would effect the
crust heating. Plausibly, the poles are then preferably
heated above the remainder of the crust, producing hot spots at
the magnetic poles.

Even if the heating in the crust is uniform\footnote{The fact that the crusts in our systems might not be fully  replaced yet by accreted matter \citep[so-call hybrid crusts;][]{wijnands2012} could significantly alter the heating reactions in the crust further complicating things.}, it is
possible that the magnetic field is not radial in the crust but has
formed a different configuration (e.g., due to the
accretion of matter in past outbursts). If it is possible to produce radial magnetic field lines at the magnetic poles yet have the lines parallel to the surface in the rest of the crust, then this
would affect the heat transport properties considerably \citep[e.g., ][]{2006A&A...457..937G,2014MNRAS.444.3198G}. If the field lines are above
the depth where heat is released, the heat can only propagate upwards to the surface at the magnetic poles. These will then be very warm compared
to the rest of the crust and may hence cause pulsed X-ray emission.  For completeness, we note that a significant amount of relevant research \citep[e.g.][]{2006A&A...451.1009P} has been performed on anisotropic thermal emission from cooling isolated neutron stars \citep[including magnetars; see the reviews of][and references therein]{2015SSRv..191..171P,2015SSRv..191..239P}.

In the cooling hypothesis, it is expected that the crust 
cools gradually until it regains thermal equilibrium with the
core. This should be observable as a steadily decreasing temperature
that causes the luminosity to slowly decay. This is exactly what we observe for both sources and therefore we prefer the cooling scenario above the other two models discussed to explain the meta-stable state (although we note that detailed modelling of direct and magnetospheric accretion has to be performed to conclusively exclude those models). The exact time scale for this
is unclear. In the low-magnetic field neutron stars, the crust cooling time
scale ranges from months to decades \citep[e.g.,][]{cackett2013_1659,degenaar2015_ter5x3}, but it is unclear how a
strong magnetic field will affect this. Further monitoring
observations with {\it Swift}/XRT are planned as well as an {\it XMM-Newton} observation for \usource\ to obtain better spectra and to search for
pulsations.

\vspace{-0.4cm}
\section*{Acknowledgements}
RW acknowledge support from a NWO Top Grant 1. ND acknowledges support from an NWO Vidi grant and an EU Marie Curie Intra-European fellowship. We thank Sergei Tsygankov and Alexander Lutovinov  for sharing an early version of their paper with us. This work made extensive use of the \swift\ public data archive. RW thanks Dany Page, Caroline D'Angelo, and Alexander Lutovino for interesting discussions.

\vspace{-0.5cm}
\bibliographystyle{mn2e}


\label{lastpage}
\end{document}